REVIEW

# PRECURSOR THAT PREDICTS THE EARTHQUAKES, TRIGGERS, AND INDICATORS

Manana Kachakhidze[1,4], Nino Kachakhidze-Murphy[1,4], Vaso Kukhianidze[2]
Giorgi Ramishvili[2], Badri Khvitia[3]

[1]Natural Hazard Scientific-Research Center, Georgian Technical University, Tbilisi, 0175, Georgia.
[2]Ilia State University, Tbilisi, 0162, Georgia,
[3]Sokhumi Institute of Physics and Technology, Tbilisi, 0186, Georgia
[4]These authors contributed equally to this work

**Correspondence**: Manana Kachakhidze kachakhidzem@gmail.com

**Abstract**
The consolidated paper presents work carried out in the sphere of earthquake problems. On the base of theoretical and experimental studies, it is shown earthquake prediction possibility. There are discussed earthquake indicators and triggering exogenous factors for the Caucasus region. Besides, because the earthquake preparation process causes anomalous changes with complex characteristics in various geophysical fields, it is given scientifically proven suggestions for the classification of earthquake precursor, indicator and triggering factors. It is offered short-term plan for future work in the earthquake prediction direction.

**Introduction**
Modern ground-based and satellite methods of observation, in the seismogenic area, during the earthquake preparation period, reveal the anomalous changes in various geophysical fields, which may accompany the earthquake preparation process, and expose themselves several months, weeks, or days prior to earthquakes. Very interesting studies have been accumulated [2,8,13,16,27] which prove the certain kind of connections between anomalous changes in these fields and earthquakes, but it turned out that in order to guide further studies in the right direction, it is necessary to classify these fields into triggers, indicators, and precursors. This classification will simplify the vague picture created by the set of anomalous geophysical fields existent during the periods of earthquake preparation, and makes it difficult to detect the true precursor necessary for high-precision earthquake prediction.

**Discussion**
**I. Earthquake precursor**

The detection of Electromagnetic radiation in the range of MHz, kHz, and ULF, existing during the formation of strong earthquakes, at the moment of an earthquake's occurrence and often after that, turned out to be so important and noteworthy that radio networks were created in Japan (Japan-Pacific Network VLF / LF) and Europe (INFREP, European Network of Electromagnetic Radiation)[2,3,14, 15, 16,17,26, 86, 93].

The regularity revealed by world-famous scientists on the basis of the research results is very significant. Namely:
1) EM emissions appear approximately several weeks before the earthquake. 2) The spectrum of electromagnetic radiation is characterized by the following sequence: MHz, kHz. 3) These emissions are accompanied by ULF radiation. 4) Before the earthquake VLF/LF electromagnetic emissions become very weak or completely disappear (so-called "silence" appears). 5) The "silence" of VLF/LF EM radiation is followed by an earthquake[13].

We raised the question as follows: if there is radiation, then there must be a radiating body, and if there is some regularity in the radiation spectrum, it must reflect the specific processes taking place in the radiating body.

In order to check this view, we have created the model of the generation of EM emissions detected prior to the earthquake[38], considering classical electrodynamics and the processes taking place in the earthquake focus.

The model allowed us to derive the formula where the frequency of electromagnetic radiation during the earthquake preparation period is analytically connected with the fault length originating in the incoming earthquake focus (1):

$$\omega = \frac{1}{\sqrt{\varepsilon\mu}} \times \frac{c}{l} = k\frac{c}{l} \quad (1)$$

where $\omega$ is the frequency of existent electromagnetic emissions, $c$ is the light speed, and $k$ is the characteristic coefficient of geological medium (it approximately equals to 1). It is possible to determine the magnitude of the expected earthquake (the first parameter necessary for earthquake prediction) by EM emissions frequency records [78]:

$$lg\ l = 0.6Ms - 2.5 \quad (2)$$

This means that a change of the geophysical field characteristic parameter (VLF/LF EM emissions frequency) is inevitably related to changes in the fault length in the focus, which allows us not only qualitatively but also quantitatively to evaluate the complete picture of an earthquake preparation from the appearance of microcracks in the focal area, up to the formation of the main fault and to a final equilibrium state.

It is logical to assume that if VLF / LF electromagnetic radiation can "work" as the precursor, it must be an exact manifestation of the mechanical stages of the avalanche-unstable geological model.

The avalanche-like unstable model of fault formation is divided into four main stages: In the first stage, when the chaotic formation without any orientation of micro-cracks takes place, can last for several months throughout the whole seismogenic area[52]. This is a reversible process since at this stage not only can form microcracks but also can appear as so-called "locked cracks". Cracks created at this stage are small. This stage, according to our model (formula 1) in the electromagnetic emissions frequency range, should be expressed by the discontinuous spectrum of electromagnetic radiation in MHz diapason [38], which is proved by the latest special scientific works [13,16,19,61].

The second stage of the avalanche-like unstable model of fault formation is an irreversible avalanche process of already somewhat oriented microstructures, which is accompanied by the inclusion of the earlier "locked" sections.

Since the crack lengths begin to increase due to the aggregation of primary small cracks, the electromagnetic radiation frequencies gradually decrease, and a continuous spectrum in MHz appears in the frequency spectrum of EM radiation[38].

The transition of the MHz emissions in kHz in the frequency spectrum of electromagnetic radiation, according to formula (1), corresponds to the very stage when the crack length already reaches about a kilometer[16,38,40,61,65].

At the third stage of the avalanche-like unstable model of fault formation, relatively large cracks combine into one main fault. This process should correspond to a gradual decrease in frequencies in kHz, which, according to formula (1), means an increase in the length of the fault in the focal area. By (2) formula, also refers to the increase in magnitudes of the expected earthquake[78].

The combination of cracks into one fault, which proceeds intensively at the final stage of earthquake preparation, will use a certain part of the energy accumulated in the focal area and, therefore, will lead to its decrease [34].

In such a situation, a period settles before an earthquake (which can last from several hours to even some days), when a fault is already formed, while an earthquake has not occurred yet, since accumulated tectonic stress is not sufficient to overcome the limitation of the solidity of the geological environment.

The system, awaiting a further "portion" of tectonic stresses, is in the so-called state of "stupor-waiting"; in principle, the process of forming main faults no longer occurs in it and, accordingly, electromagnetic emissions do not occur, which has been proven by experiments[18,19,61].

This process is expressed correspondingly in the electromagnetic emissions spectrum: some hours before the earthquake (up to 2 days) in the spectrum the emissions interruption or "electromagnetic emissions silence" is observed [61]. Of course, we should expect the renewal of electromagnetic emissions exactly before the earthquake (the fourth stage of the fault formation process).

At the final stage of earthquake preparation, a very short time is required to replenish the critical reserve of tectonic stresses and the origination of the main fault, which means that when monitoring electromagnetic radiation, for determining the time of occurrence of an incoming earthquake (the second characteristic parameter for prediction) the moment of interruption of the EM radiation spectrum is important.

Thus, good conformity and synthesis of our model and avalanche-like unstable model of fault formation are evident.

In addition, for determining the epicenter (the third characteristic prediction parameter) VLF/LF EM radiation is useful. Namely, from the points selected around the receiver, where VLF/LF EM emissions are fixed, by the Direction-finding method, to define the incoming earthquake epicenter is possible.

Thus, it can be emphasized that VLF/LF EM emissions turned out to be a unique precursor since they make it possible to simultaneously determine the magnitude of an earthquake, the epicenter, and the time of occurrence of inland moderate and strong incoming earthquakes. This is the first and only precursor that describes the process of formation of faults in the focus of an incoming earthquake and numerically calculates the length (magnitude) of the fault at any moment of monitoring.

It is worth underlying that no reliable criterion in seismology yet, can distinguish the strong foreshocks from the mainshock. This issue can be solved with rather a high accuracy on the basis of a theoretical model[38]: if after any shock electromagnetic emissions still continue to exist and the frequency data still tend to decrease, it means that the process of fault formation is not completed yet and we have to wait for the mainshock. It is on the contrary in the case of aftershocks.

In the next step, our interest was to check our theoretical study with the retrospective continuous data and to "predict" an already occurred earthquake.

We worked out 73 days of continuous INFREP data for the Crete earthquake with M= 5.6 (25/05 / 2016, 08:36:13 UTC)[39].

INFREP network fixes every minute amplitudes of 10 different baseline frequencies of VLF/LF electromagnetic radiation in diapason 20 270 Hz - 270 000 Hz.

Based on our studies, if any frequency channel actually reflects the earthquake preparation, the avalanche-like unstable geological process should be reflected in the frequency data of this channel. We calculated in every minute the corresponding lengths of cracks in relation to the base lengths of all corresponding frequency channels (as a percentage).

It was found out two active channels (37 500 Hz) and (45 900 Hz) as the average daily value of the crack lengths was the maximum for them. This means that from the discussed 10 channels, the earthquake preparation process only in two of the above-mentioned frequencies had been described. In this case, according to the above-given formula (2), the magnitude of the incoming earthquake should been between 5.5 and 5.7 (Crete earthquake magnitude is really estimated as M= 5.6).

In addition, we found that the diurnal periodic variations of electromagnetic radiation, with the exception of the channel (37500 Hz), are clearly expressed in all channels. Such variations in the (37 500 Hz) channel are recorded too, but till some period, namely up to 02.05.2016, after which the anomalous process starts, indicating that the avalanche–unstable process of fault formation has already begun (Fig. 1).

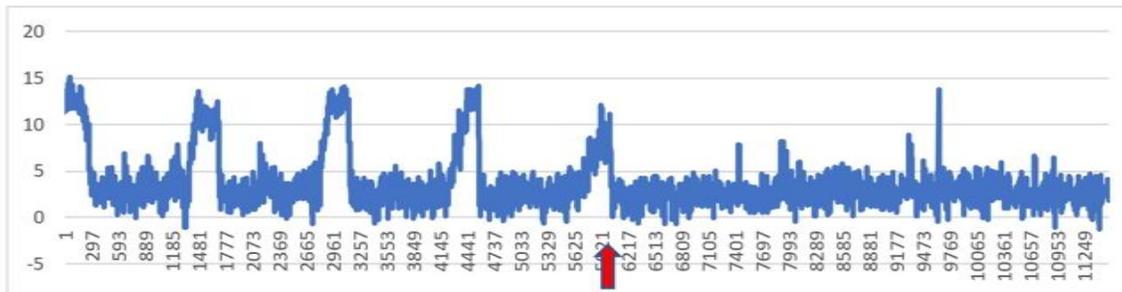

Fig. 1.

Figure 1. The initial moment of the avalanche-like unstable process of fault formation is noted by the arrow.

Since only the (37 500 Hz) frequency channel meets both conditions: the average daily value of the crack's lengths is maximal for this channel and an avalanche process appears only on it. Therefore, to predict the earthquake, we must rely only on the data of this channel.

During the analyses, it was found that 19 days before the earthquake, an avalanche process of fault formation occurred in the frequency channel (37 500 Hz).

In order to analyze the earthquake preparation whole process, we elaborated the daily averaged frequencies by using the average square deviation method and calculated $\overline{x} \pm \sigma$ values (Fig.2).

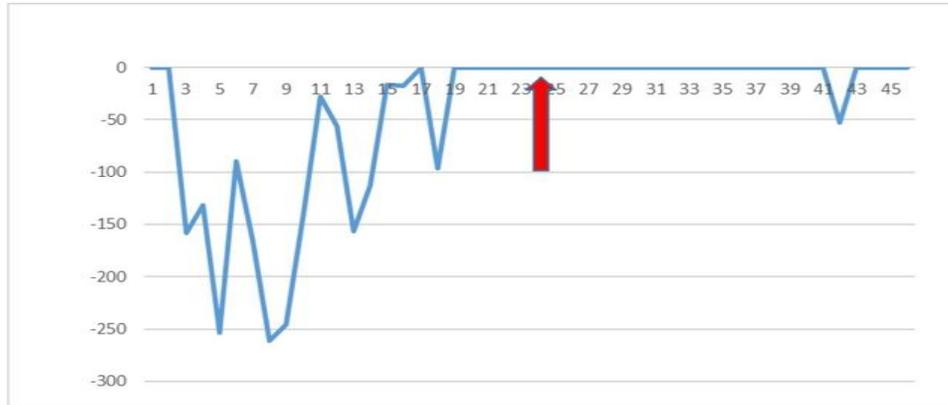

Fig. 2.

Figure 2. The avalanche process of fault formation and EM emissions "silence" period before the earthquake (the arrow noted the Earthquake occurring moment).

Such elaboration of the data manifests the EM emission's "silence" starting moment prior to an earthquake. It was equal to about two days.

The fact that retrospective analysis of data from real earthquakes makes it possible to fix the exact moments of the starting of the avalanche process of fault formation and of the "silence" of EM radiation means that in the case of monitoring EM emissions, it is possible to observe the entire process of preparing earthquakes and predict earthquakes.

It is clear that when the tectonic stress increases in any region's certain zone different physical-chemical fields will change in this area and most of them will be observed as anomalies. The features of these fields in connection with earthquakes are revealed in the fact that they usually, only qualitatively express the processes in the focal area. Based on them it is not a possible simultaneous determination of all three parameters necessary for incoming earthquake prediction (magnitude, epicenter, and time of occurring). It means, that these fields really have only indicative features in the earthquake preparation process.

It is obvious also, that the process of fault formation without fail must be reflected in the abnormal changes in the certain geophysical field and this phenomenon should be recorded prior to the earthquake. The prediction of the incoming earthquake will be possible only by analyzing an anomalous change in such a field and as it is mentioned above, this field must be referred to as an earthquake true precursor (or simply a "precursor").

In case of weak earthquakes, we have to wait for the electromagnetic emissions in high frequency diapason but these waves attenuate rapidly, and observing them on the earth's surface is difficult.

Our research allows us to think about the possibility of artificial removal of tectonic stresses with using seismic and electrical influences, about which relevant articles by Russian scientists were published at the end of the last century[57,59,74]. Based on their own theoretical and experimental research they wrote: "today, at the modern stage of scientific development, the task of gradually easing the energy accumulated in the focus of a large earthquake for the reduction of tectonic stress is completely real. To achieve this goal, it was necessary to find a place where nature already has prepared for a catastrophic earthquake. In other words, it is necessary to solve the problem of forecasting an earthquake" [57,59,74].

Today, because the precursor has already been discovered, a VLF/LF EM emissions mobile receiver has been created, and in the case of creating a VLF/LF EM emissions mobile network, in any s/a country, it is possible to make the short-term (operational) prediction of strong earthquakes, it appeared opportunity to carry out a controlled discharge of tectonic energy and reduce the seismic hazard by using special, reasonably organized impacts on the earthquake focus [57,58,74].

## II. Earthquakes' triggers
II.1. **Tides as earthquake triggers**

It is known the exceptionally high sensitivity of the processes occurring in the focus of impending earthquake to external natural and artificial influences: for example, changes in the speed of the Earth's rotation, terrestrial tides, solar activity, weather events, earthquakes, creation of large reservoirs, development of deposits of oil, gas, solid minerals, pumping of liquid industrial waste, conducting underground nuclear tests, sounding the earth's crust with powerful electrical impulses, etc.

Any above-mentioned factors and among them certain exogenous (cosmic) triggering factors may exist in every seismoactive (s/a) region. They can play a triggering role and correct the earthquake's occurrence moment in the case when the focal area is on the limit of the critical value of the geological environment [12,24,32,49,58,67,72,73,89,92].

The task of determining the exogenous action on the region from the main celestial bodies for regions at any point and any moment of the time can be solved by taking into account their geographic coordinates[42,43].

As known the different total tectonic stress acts in the different regions because of plates movements particularities. In Caucasus tectonic stress is caused by advancement of the Arabian lithosphere plate to the North and its reapproachment[33,37].

Basing on the geological and geophysical data we can pick out:
1. Interzonal deep faults of the Caucasian orientation, which divide the main tectonic units of the region.
2. Interzonal longitudinal and transversal faults.
3. Transzonal faults of the Anticaucasian and submeridional trend.

At present these faults determine heterogeneity of the Caucasian stress field, which is mainly characterized by the shear stress, directed diagonally towards the main structures and finally-by the tensile stress, directed parallel with the main structures.

Because, it is not possible to determine value of the force caused by plate's movement and the task of N – body is not solved analytically yet, we investigated earthquakes exogenous triggering factors by numerical methods.

It is known that any planet is not only under the Sun attraction but under other bodies of the Solar system also. As the planets of the Solar system have a mass many times less than the Sun, the Sun and the Moon are considered as the "main bodies" of perturbation.

We consider earthquakes occurring moments connection with the Sun, the Moon attractive and gravity perturbation components of tide[37].

We have discussed all 393 Caucasus earthquakes of 1900-1992 years with M $\geq$ 4.5 (we have to underline that here, and below, where data of geophysical fields are elaborated, we were limited by data up to 1992 because, after 1992, due to certain objective reasons, getting reliable, systematic geophysical data in Georgia, was complicated).

In order to detail analysis of task, the Moon, the Sun and their total perturbation directions were calculated and inflicted on the Caucasus main faults map for each earthquake separately (For example: Fig.3,4,5).

We considered frequencies of recurrence of azimuths in intervals of tides possible directions.

Therefore, we had revealed triggering factors, the meanings of the azimuths of the Sun and Moon at the moment of the occurrence of the significant quantity of earthquakes, and estimated validity of the cases.

It was found out, that the sectorial effect is obviously shown in the Moon's gravity perturbations and in the total gravity perturbations of the Sun and the Moon (Table 1).
Table 1.

Table 1. Azimuths of the perturbation forces

| Triggering factor | Azimuth of the force direction | Validity of the phenomena |
|---|---|---|
| Perturbation from the Sun | 193°–206° | 98% |
| Perturbation from the Moon | 295°–310° | 99.8% |
| Total perturbation | 290°–307° | 91.5% |

It should be noted that the horizontal component of the gravitational field of lunisolar tides, for our latitude, is never directed to the north, i.e. it is never directed within a sector which azimuth is between $323^0$-310°.

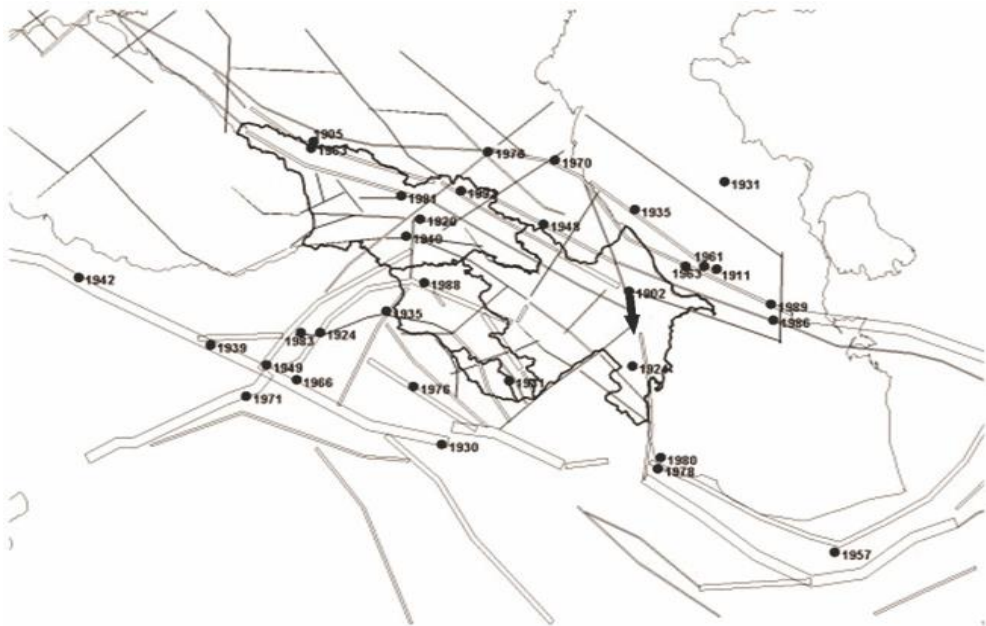

Fig. 3
Figure 3. the Moon tide direction at Caucasus 1902 earthquake occurring moment;

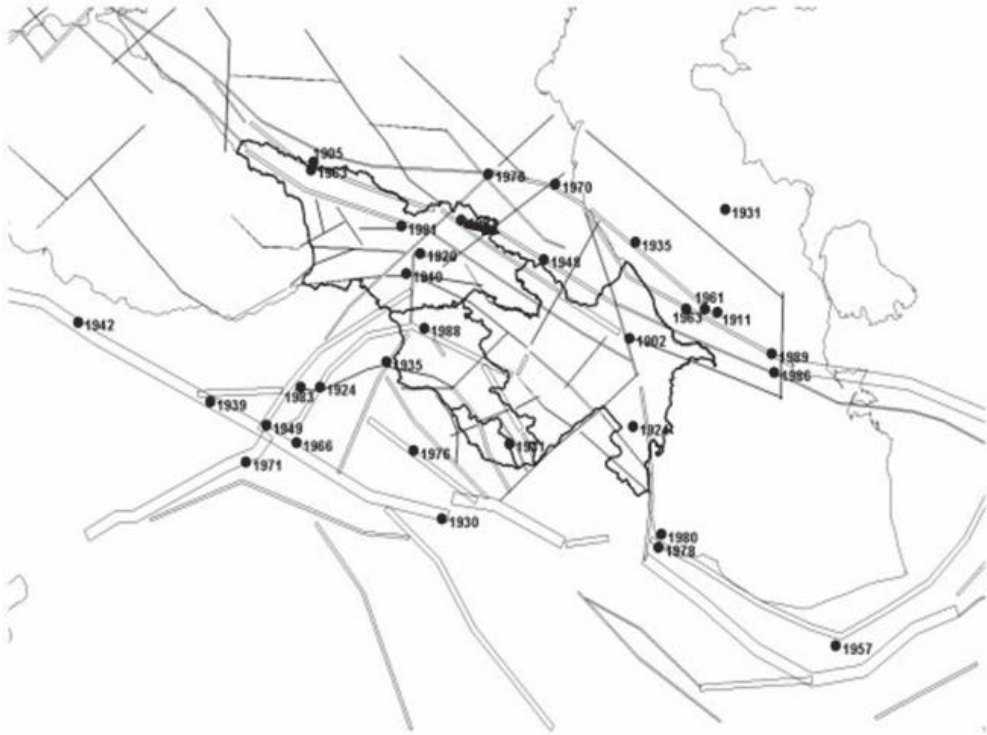

Fig. 4

Figure 4. the Sun tide direction at Caucasus 1992 earthquake occurring moment

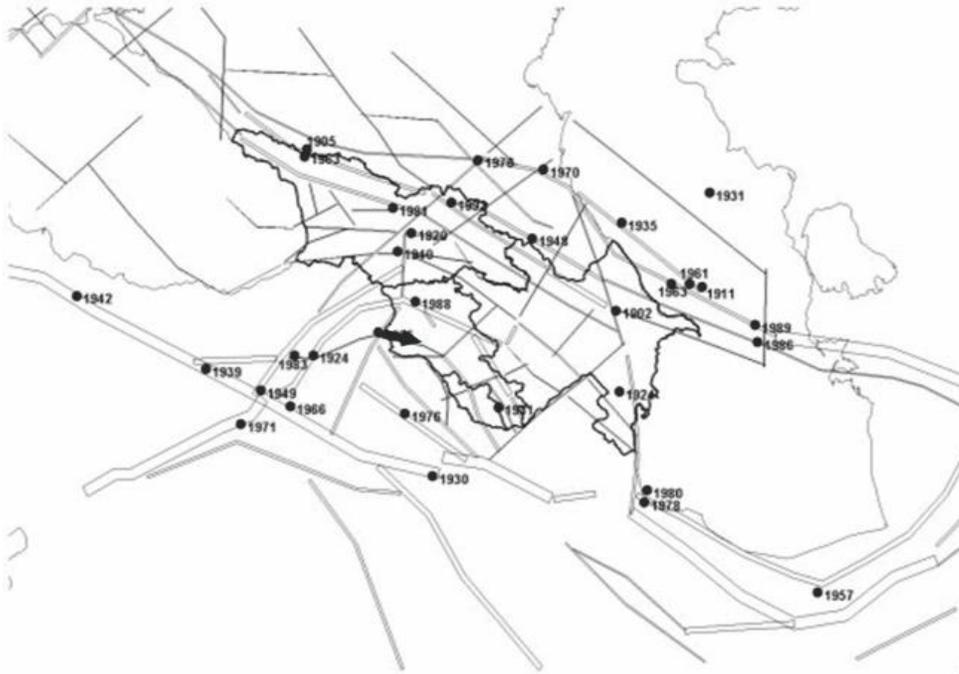

Fig. 5

Figure 5. total tide direction at Caucasus 1935 earthquake occurring moment (tide directions at the occurring moments of the earthquakes are marked by arrows).

Comparing the obtained results with the tectonic map of the Caucasus, we can conclude that the largest number of earthquakes in the Caucasus with M≥4.5 occurs when:
1. The gravitational field and solar perturbation are perpendicular to the plane of the Russian Platform (i.e., perpendicular to the fault along the Caucasus ridge) and directed from North to South, in particular, creating angles of 20°–30° with the $40^0$ meridians.
2. Lunar and total disturbances coincide with the direction of tectonic stress from approximately southeast to northwest (along the mentioned fault in the Caucasus seismically active region) and are known as tensile stresses. They create $60^0$ angle with $40^0$ meridian.

As for large earthquakes in Greece, the obtained results allow us to draw the following conclusion: earthquakes occur more often during moonrise or moonset, i.e. during the outflow (lunar tide), and strong earthquakes during the inflow.

Based on our study we concluded that in the case when similar works are carried out in any s/a region that can cause a trigger effect, we must take into account the regional tectonic stress situation, above described distributions of the Sun and Moon on the sky, and relevant time intervals as the periods of increasing danger of strong earthquakes occurring in the region.

Obviously, the stress acted on the Earth from the Sun and Moon, and, in our case, on the Caucasus, can play a role of the triggering factor only in case its direction helps to increase tectonic stress in this region. Of course, at the same time, these additional stresses affect neighboring regions (for example, Turkey and the Arabian Peninsula), but their tectonic stresses have other directions. Because of this the force acted on the Caucasus region, from the Sun and Moon, cannot play the role of the triggering factor for Turkey or the Arabian Peninsula. On the contrary, the exogenous stress added to Turkey's tectonic stress can appear as an earthquake initiative factor for Turkey and cannot play the role of the triggering effect for the Caucasus and Arabian Peninsula earthquakes.

**III. Earthquakes indicators**
**III.1. Relatively weak earthquakes of the Caucasus as strong earthquake indicators**

Many interesting papers are devoted to the seismicity changes regarding major earthquakes. There are revealed some seismic quiescence, and some seismic activations as well before comparatively strong earthquakes which have to been caused by regional geological particularities [9,10,56, 62,85,91].

From the geological and geomorphologic viewpoints, Javakheti is one of the most complex and seismic active parts of southern Georgia (Caucasus)[35]. The majority (55%) of earthquakes observed throughout the Caucasus occur here. The question arises as to whether Javakheti reflects the regional geophysical processes. To find out if Javakheti is "sensitive" to strong earthquakes, we considered 695 Javakheti relatively weak earthquakes (2.5 ≤M < 6.0) observed for 1961–1992 with regard to occurrence moments of strong (M ≥ 6.0) 16 earthquakes. The change over time in the monthly number of weak earthquakes considered in relation to strong ones gave a rather vague picture.

To separate any possible anomalies from obtained distribution, we used the method of the Creeping Mean [23]. We smoothed data series, from 2 to 20 months in length, with a lag of one month until getting of the sharp picture. The 9-month series were found the most appropriate for analyzing (we have to note that 9-month series was chosen empirically as the best signal/noise ratio among other variants).

In the next step we have applied a qualitative or Sign-Changing Summation Method [23] which is quite convenient even in case when the amplitude of disturbances equals or exceeds the amplitude of "useful" anomalies.

Analysis of the results, by using a gradual approximation, makes possible to delineate the certain zone with margins from (+53.1) to (−51.8). The "useful" anomalies, observed beyond the zone, gave evidence of weak earthquakes "sensibility" to incoming strong earthquakes (Fig.6).

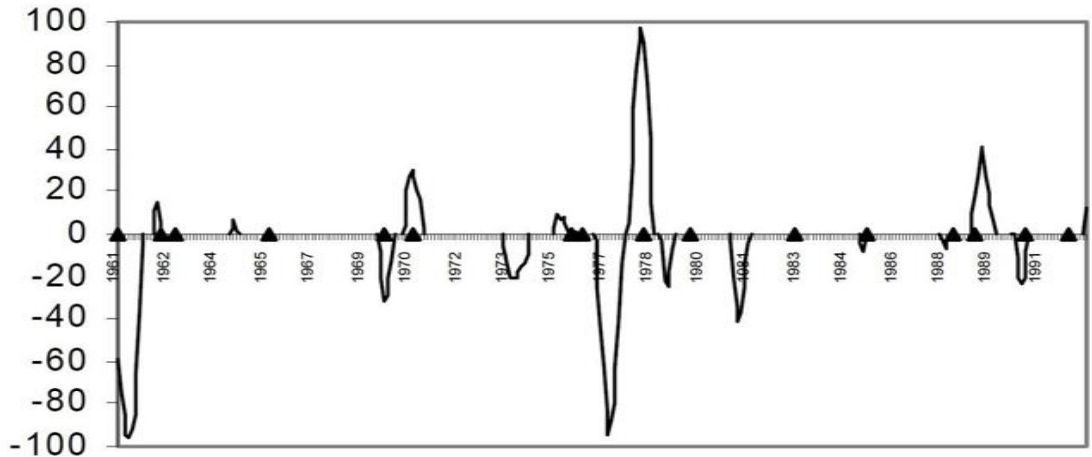

Figure 6. Graph of anomalous distribution of Javakheti weak earthquakes towards Caucasus strong ones

By obtained graph, the minimal positive anomaly is equal to (7), and negative anomaly is (-8). 4 earthquakes, out of all strong considered earthquakes, occurred during anomaly and all the rest earthquakes occurred by following: anomaly, aseismic periods and earthquake. The average continuance of the aseismic period is estimated as 5.3 month (we underline that here and below all numerical values are graphical and they should not be relevant to real values. They are helpful in monitoring process).

At the next stage, we considered a similar task for the whole Caucasus region earthquakes.

Due to the fact that $2.5 \leq M < 4.5$ earthquakes were not reliably registered before 1961, we divided the task into two parts. We considered:

1. $4.5 \leq M < 6.0$ earthquakes for 1900-1992 and
2. $2.5 \leq M < 6.0$ earthquakes for 1961-1992.

33 strong ($M \geq 6.0$) and 360 earthquakes with $4.5 \leq M < 6.0$ occurred in the Caucasus in the period 1900-1992. After determination of $4.5 \leq M < 6.0$ earthquakes number in each month, we used the method of the Creeping Mean, and with the same, above-pointed opinion, stopped at the 9-month series for smoothing.

After using of the qualitative method of separating the "useful" anomaly, and the gradual approximation method (we cut (+35) and (-43) zones), was clearly revealed the anomalous change of Caucasus $4.5 \leq M \leq 5.9$ earthquake number in time with respect to strong earthquakes occurring moments.

The minimal positive anomaly value is equal to (28), and negative – (-131). 11 earthquakes occurred by following: anomaly, aseismic period and earthquake, and other strong earthquakes occurred during anomalies. The average duration of the aseismic periods is equal to 8.9 month.

7064 earthquakes occurred in the Caucasus for 1961-1992 with $2.5 \leq M < 6.0$. Similar to the two above tasks, we estimated number of $2.5 \leq M < 6.0$ earthquakes by month and after we processing data by all above-pointed methods (by gradual approximation method we cut (+115) and (-332) zones).

The minimal positive anomaly is estimated as (83), and minimal negative anomaly as (-45). All strong earthquakes occurred by the following sequence: anomaly, aseismic period and earthquake. The average duration of the aseismic period is 6.7 month.

In case of this task, it found out that, before only one earthquake was not revealed an anomalous distribution of $2.5 \leq M < 6.0$ earthquakes, and also only one anomaly was detected that was not followed by a strong earthquake (a so-called "false alarm").

We may underline that although in the discussed tasks we studied the different (695, 360, and 7064) numbers of earthquakes with different magnitudes, a general regularity was revealed for the Caucasus region.

In order to isolate so-called "useful" anomalies, in all three cases, the 9-month smoothing method was selected as the best among the different variations and final results are obtained by using of the same qualitative method of anomaly isolation.

In the case of 31 out of 33 strong earthquakes an anomalous change in the number of earthquakes of different intensities (M<6.0) with respect to the occurrence moment of the strong earthquake was detected. In majority cases the following sequence is maintained: anomaly, aseismic period and earthquake.

Based on the generalization of the obtained results, it can be concluded that a special seismicity of earthquakes with $2.5 \leq M < 6.0$ and $4.5 \leq M < 6.0$ in the Javakheti, as well as in the entire Caucasus, reflects such a global process as the preparation of a strong (M >6.0) earthquakes in the region, and the region, as a unified tense system, is "sensitive" to the preparation of strong earthquakes, that is, the $2.5 \leq M < 6.0$ and $4.5 \leq M < 6.0$ earthquakes manifest themselves as "regional foreshocks".

We think that such specific seismicity of relatively weak earthquakes can be considered as an indicator of strong earthquakes.

### III. 2. Specific variations of the atmospheric electric field potential gradient before the Caucasus $M \geq 5$ earthquakes

Much attention is paid to the parameters of the atmosphere, for instance, such as it is an atmospheric electric field. Abnormal changes of the electric field potential gradient (EFPG) have been observed in the near-ground air before earthquakes in different regions of the world [31,68,69,70]. Authors have their opinions about the cause of the existence of this anomaly but there is no mutual agreement among them.

Our aim was to investigate a possible existence of "seismic share" in anomalous disturbances of atmospheric electric field. In order to study this task, we have considered Dusheti (Georgia, Caucasus) observatory's EFPG records by hours from 1956 up to 1992 with respect to Caucasus earthquakes 41 (M ≥ 5.0) occurrence moments[36].

Investigation was carried out for 11-day EFPG data before earthquakes.

General picture of EFPG distribution did not show hopeful imagination for indicatory character of the considered parameter.

Knowing the nature of atmosphere's electric field, first of all we should pay attention to its seasonal variations. Because of it, all data of EFPG were classified according to months and average quantity $\bar{x}$ of background value was calculated for each month. In the next step we computed corresponding s dispersion, $\sigma$ deviation and $\bar{x} \pm \sigma$, $\bar{x} \pm 2\sigma$ and $\bar{x} \pm 3\sigma$ significances for 11–day EFPG data before all considered earthquakes.

Anomalies, which are received by cutting off $\bar{x} \pm \sigma$, $\bar{x} \pm 2\sigma$ and $\bar{x} \pm 3\sigma$ quantities from the EFPG data, were named as "usual" anomalies of I, II and III types correspondingly.

Of course, the problem of meteorological parameters' influence on EFPG quantities should have been considered during data analysis.

As known, the average meaning of EFPG for Dusheti Observatory is 84 V/m in normal weather conditions (low clouds amount is not more than 4, wind velocity is 4 m/sec or less, and there are no precipitations, thunders, and lightings).

In order to identify "clear" anomalies connected only with earthquakes, we have considered 7 basic meteorological parameters: 1. low clouds, 2. CB clouds, 3. precipitations, 4. wind, 5. air temperature, 6. relative humidity and 7. absolute humidity. Besides these parameters we have considered cases of thunders and lightings. All disturbances of atmospheric EFPG, which were provoked by thunders and lightings, were disregarded.

In order to exclude inter overlapping influence of meteorological parameters we have made attempts to estimate EFPG anomalous disturbances which were caused by above mentioned 7 basic meteorological parameters. Each meteorological parameter was divided into three groups: high (strong), average (moderate) and low (weak).

So we have received 21 "sorted out" meteorological parameters instead of 7.

On the basis of new 21 meteorological parameters we have created so-called "weather combinations" in the following way: in case of fixed quantity of one meteorological parameter (for example, weak wind) we have considered such weather combinations when all other meteorological parameters are strong or high.

On the next stage, in case of the same value of the same parameter (weak wind), we have considered new weather combination, when the rest of parameters, except one (for example, precipitation) are high or strong, but precipitation is moderate.

Finally, we have received 2187 weather combination where none of possible variants is omitted.

The creation of so-called weather combinations enabled us to estimate corresponding average quantities of EFPG for every concrete weather situation and only then we have returned to previously identified "usual" three type anomalies and started to filter out them.

Due to filtration we have received new anomalies, which were named by us as "clear" anomalies of I, II and III type.

To our opinion all classes of "clear" anomalies are important, but the "clear" anomaly of the III type is the most significant because all randomness and disturbances due to weather effects are excluded. These anomalies in the majority of cases, manifest themselves in 4 day's period before M$\geq$ 5.0 earthquakes. Quantity of III type "clear" anomaly varies from (-152.1) V/m to (277.3) V/m (significant values comparing to average annual quantity of atmospheric EFPG for Dusheti observatory). Duration of III type "clear" anomalies vary from 40 to 90 minutes and are revealed in 29 cases (71 %) out of 41 discussed earthquakes of the Caucasus.

We have tried to reveal a correlation from one side, between values of EFPG anomalies and earthquake magnitudes values, and from another side, between values of EFPG anomalies and epicentral distances (from Dusheti Observatory), but we have not revealed any interesting relations for the Caucasian region.

Therefore, the "clear" anomaly of the III type is recognized as an earthquake indicator.

### III.3. A study of atmospheric pressure variations that existed prior to an earthquakes

In scientific works, among the meteorological processes occurring during the period of earthquake preparation, variations in the atmospheric baric field are often considered as the initiating factor of an earthquake [30,50,71]. To confirm this view, as an example, it is cited the change in atmospheric pressure relative to the Chile earthquake (1960, 21 May, M = 8.5) occurring time and is concluded that in that case there were strong disturbances of the atmospheric bar field. Similar effects are observed during catastrophic earthquakes in different regions [1,64,87,88].

In order to reveal the anomalous effects of atmospheric pressure relative to the earthquake occurrence moment and the epicenter, we examined 40 earthquakes (M ≥4.5) in the Caucasus region during 1966-1992 years.

Based on the atmospheric pressure isobaric maps, the time-variations of the 7-day atmospheric pressure data for 5 days before and 2 days after the earthquake was evaluated.

Since the epicenters of earthquakes are located at different heights from the sea level, for the accuracy of the atmospheric pressure values, each record was brought to the corresponding pressure value regard to epicenters heights (as a result, we got the so-called pressure reduced values).

All considered 7-day curves of changes in atmospheric pressure relative to the epicenters and the occurring moments of considered earthquakes were very indeterminate for analysis. Therefore, for revealing possible anomalous change we used Sign-Changing Summation Method [23].

As a result, we got a different curves, one example of which is presented in Fig. 7. Based on the analysis of the obtained curves for all earthquakes, out of 40 considered earthquakes, depending on the epicenters of earthquakes and the moments of their occurrence, a sharp anomalous change in atmospheric pressure values was found in cases of 34 earthquakes. For each earthquake, we estimated the quantity of the atmospheric pressure change by calculating the ratio of the atmospheric pressure gradient to its average value corresponding to the specific 7-day period under consideration.

It was found that for the cases of only two earthquakes, the quantity of atmospheric pressure change is $10^{-3}$, and in all other 32 cases it equals to $10^{-4}$.

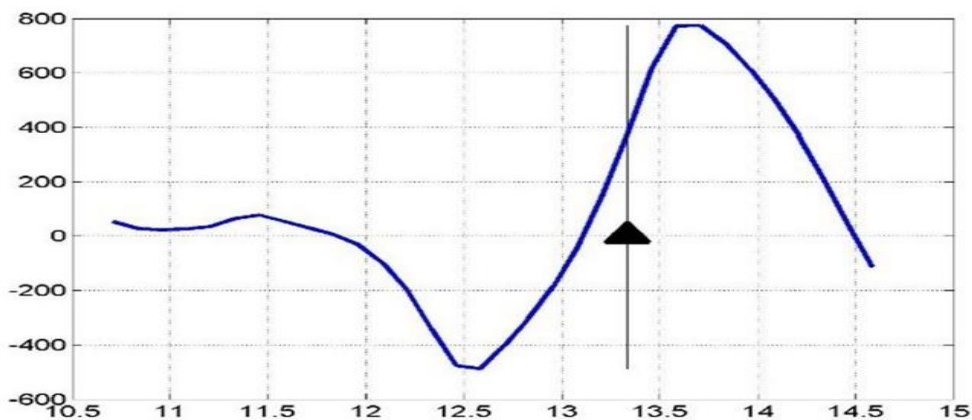

Figure 7. The curve obtained as a result of the processing of atmospheric pressure data with respect to the 1986.05.13 earthquake occurring moment by Sign-Changing Summation Method.

zone, causes the generation of electro-telluric variations. In this case, the telluric current obtains a vertical direction from the earth to the ionosphere, causing the polarization of the magnetic field, and besides the earth's surface gets the positive potential. Such allocation of fields coincides exactly with the distribution of the same fields in bad weather conditions which is companied by atmospheric pressure changing [41].

Thus, variations of atmospheric pressure are related to changes in the telluric field, which occurs in all regions during both earthquake preparation periods and bad weather. Therefore, we think, that the atmospheric pressure anomalies, revealed by us, may be considered as an earthquake indicator. Taking into account works mentioned above, it is not excluded that the anomalous changes in atmospheric pressure towards earthquake occurrence moment, have a dual, indicatory and trigger nature.

### III. 4. Earthquakes and related anomalous ULF electromagnetic radiation

The dynamic processes in the earthquake preparation zones can produce current systems of different kinds [44,53] which can be local sources for electromagnetic waves at different frequencies, including ULF. These waves can propagate through the crust and reach the earth's surface, unlike high-frequency waves [7,25,29,48,82].

Thus, in ground-based observations, we could expect some ULF signals of seismic origin observed in both geoelectric and geomagnetic fields [44,79,80]. Their disturbances in the preparation zone were considered a factor of such importance that in a number of works it was proposed to use telluric variations as a short-term precursor of strong earthquakes [46,76,81,82,83].

However, research does not prove this fact.

Since tectonic stress basically "works" for the formation of the main fault in the earthquake focus, in other parts of the seismogenic area, it can no longer form the necessary conditions for the occurrence of an earthquake. However, tectonic stress can cause perturbations of the geophysical fields in the seismogenic area.

As a result of the growth of tectonic stress, heterogeneity appears in earthquake preparation areas [21], and like to "Frankel's generator" this segment of the earth's crust will have inductive polarization [20,47,53]. Generally, polarization charge should be perturbated over some surfaces [90]. Indeed, experimentally has been proved that electric dipoles appear on their surface [16,17,19,22].

This polarization charge takes part in two different processes:

**a)** If microcracks coalesce in the form of any size rupture nucleus (including the main fault) and in the fault plane there are changes in the specific electrical resistance of the rocks, that is there are inclusions of high electric conductivity, it is not excluded that the fault and the layer on which the polarization charge is distributed, might be locked by a vertical electric field like a double-wire conduction layer, which form an oscillating contour-like structure.

At the certain stage of earthquake preparation, the avalanche-like process of fault formation begins and ends with an earthquake, accompanied by emissions of electromagnetic waves of VLF/LF frequency. The value of the electromagnetic wave frequency emitted at this time depends on the length of the fault. In the case of small cracks, the electromagnetic radiation will be of the order of MHz, and in the case of cracks of the order of *km,* it goes into kHz [38,40].

**b)** In areas where tectonic stress is not able to form a crack, and because, in general, rock density increases with depth, in case of the same tectonic stress effect, more inhomogeneities appear in the upper rocks with lower density compared to the lower rocks, that is, in the upper rocks more polarization charges are generated compared to the lower ones.

However, since the polarization charge must be distributed over some surface, each of these surfaces will be approximately equipotential [90].

The work done during the movement of the charge $q_0$ on the $\Delta l$ element of the equipotential surface is equal [84]:
$$\Delta A = q_0(\varphi_1 - \varphi_2) \quad (3)$$
This work can also be represented by means of field tension:
$$\Delta A = q_0 E \Delta l \cos\alpha \quad (4)$$
According to formulas (3) and (4):
$$q_0 E \Delta l \cos\alpha = 0$$
and since $q \neq 0$; $E \neq 0$; $\Delta l \neq 0$, therefore $\cos\alpha = 0$ i.e. $\alpha = \frac{\pi}{2}$.

Since the lines of force in any electrostatic field are perpendicular to the equipotential surfaces and the voltages are summed vectorially, the total direction of the voltages of the electric field created by the polarization charges of these inhomogeneities will be the same.

It is known that telluric electric field at any point change its direction and magnitude all the time. At the same time at perturbation telluric field becomes linearly (or plainly) polarized. Therefore, one has to assume that under conditions of increasing tectonic stresses, the telluric field is predominantly perturbed and polarized. If we exclude the effects that external factors can cause, the changes in this field in a given area will uniquely depend on the changes in tectonic stresses.

Thanks to contact of solid or gaseous phases existing between the earth and atmosphere, diffusion of electrons and ions and ion adsorption take place, which conditions the creation of a stable electric layer (dipole layer) on the contact. In this layer, the electric field, supported by factors conditioned by earthquake preparation, can be called an "additional" electric field and can be marked as *En* (5). The electric field potential at the separating border of these two mediums suffers discontinuation, which equals contact "additional" electric field strength:
$$\varepsilon^{add} = \int_1^2 E_n^{add} dn \quad (5)$$
where 1 and 2 points are located on both near boundary of the contact surface. The fact of the mentioned field discontinuation will be expressed in all geophysical phenomena connected with the "additional" field [45] (Fig. 8):

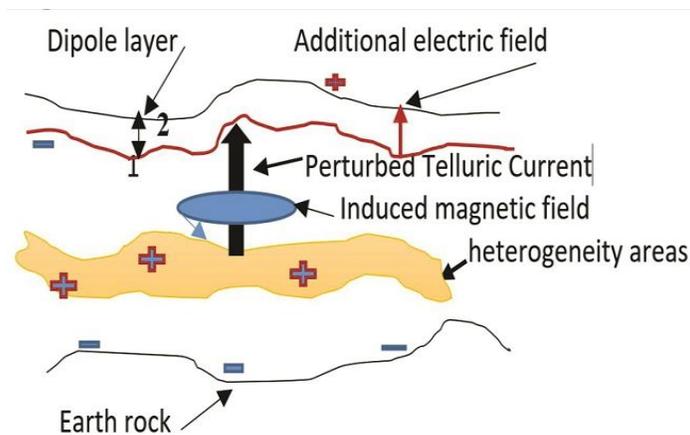

Figure 8. Perturbed Telluric current.

Since the charges move in the direction of the field line, at this time the electric current generated in the rock, is caused by the changes in mobility of charged dislocations [77] and/or point defects [21].Obviously, under these conditions, the direction of the telluric current, will be polarized and directed vertically in the direction of the Earth-Ionosphere, which also is observed in nature [63].

Telluric current is practically a synonym for geoelectric potential difference and preseismic telluric current signals, called seismic electric signals (SES) [60].

As this current is associated with the displacement of charges, it releases heat (changing the orientation of the dipoles causes heat release) [84], so the temperature in this area will increase, which is also confirmed by the experiment [66,75].

The telluric current will generate a magnetic field by induction, and as a result of the current changes in the Earth's crust the anomalous magnetic field variations start (which, according to Maxwell equations, should be accompanied by a strong SES activity)[6].

According to Maxwell's theory, in the case of plane electromagnetic waves, a change in the electric field directed along the axis OZ leads to the generation of a magnetic field directed along the axis OY(8).

$$\varepsilon\varepsilon_o \frac{\partial E_z}{\partial t} = \frac{\partial H_y}{\partial x} \quad (6)$$

i.e. the induced magnetic field will also be polarized due to the polarization of the its OY component (Fig. 8), which is observed in nature during earthquakes[4,5,8 ,10,27,28,48].

Unlike the electric field, which suffers discontinuation at the Earth-Atmosphere boundary, the induced magnetic field propagates in the Earth-Ionosphere zone, creating a polarized eddy electric field in the atmosphere.

The perturbated magnetotelluric field should manifest itself already at the first stage of the main fault formation. This fact was confirmed by laboratory and field observation [54,60].

It should be taken into consideration that magneto-telluric field perturbation will take place not only during the period preceding the earthquake but also after it too, till tectonic stress accumulated in the focal area is released completely [55].

Thus, during the preparation, occurrence, and subsequent period of the earthquake, including the aftershocks attenuation, changes in the telluric currents are directly related to changes in tectonic stress, so this field should be considered only as an indicator of an earthquake.

**Results:**

Anomalous fields, observed in the seismogenic area during the earthquake formative period, must be divided into three groups: earthquake triggering factors, earthquake indicators, and earthquake precursors, which can be defined as follows:

1. The physical field that exists independently of the earthquake preparation process should be considered as an earthquake trigger (natural or man-made) if at the final stage of earthquake preparation when tectonic stress reaches the limit of the solidity of the geological medium, can affect the tectonic stress and correct the time of earthquake occurrence (for example, tides, in some cases changing of atmospheric pressure, reciprocal - influences of earthquakes, water level rapid change in reservoirs, etc.).

2. An indicator of earthquakes can be considered a physical field that, due to the earthquake formative process, undergoes abnormal changes since a certain period of earthquake preparation but it should be emphasized that the origin of the earthquake indicator is not caused by the direct geological process of the main fault formation. It is caused by perturbations of geophysical fields in seismogenic (mostly focal) areas (e.g. earth magnetic field anomalies, changes of atmospheric electric field potential gradient, telluric field anomalies, gases emissions from rocks, TEC anomalies and etc.).

Therefore, "earthquake indicators" can be called those anomalous changes of geophysical fields, which are caused by the side effects of earthquake preparation process, cannot analytically describe the process of main fault formation, that is, these fields are not useful to predict the earthquake.

3. The true precursor (VLF/LF EM emissions) of an earthquake (simply, a precursor of an earthquake) is a physical field, which is caused by the process of the main fault formation in the focal area of the impending earthquake at the expense of the arising and coalescence of the cracks. By reflecting this process on the Earth's surface, an anomalous change in the parameter of this field (or parameters) makes it possible simultaneously to predict the magnitude, occurrence time and location of an impending earthquake with high accuracy. Our study found that VLF/LF EM emissions are responsible for earthquake short-term prediction.

**Global Implications and Future Perspectives**
To search for natural disasters such as earthquakes is one of the prime goals of the Natural Sciences. Besides taking hundreds and thousands of human lives, devastating earthquakes can have a very pernicious effect on infrastructures like, for instance, mines, reservoirs, atomic power stations (APSs), etc., which in turn may lead to ecological disasters with long-lasting consequences for the affected region. Many secondary effects induced by earthquakes (landslides, debris flows, avalanches, etc.) bring extensive damage to the seismic active countries. The rate of risks associated with these hazards increases every year due to the appearance of new complicated technological construction. Studies of earthquake problems in the world were especially intensified in the second half of the past century since alongside theoretical studies it became possible to carry out high-level laboratory and satellite experiments. Thanks to them in the earthquake preparation process various anomalous changes in geophysical fields have been revealed in the lithosphere as well as in the atmosphere and ionosphere. Among the anomalous geophysical phenomena preceding earthquakes, specific attention is attributed to the earth's electromagnetic emissions before earthquakes. In recent decades, the networks for collecting VLF/LF radio signals have been organized in many seismically active countries of the world. The role of electromagnetic emissions networks is most important because they enriched science with invaluable information and made the problem study far more wide-scale.

Satellite and ground-based observations proved the presence of VLF/LF and ULF electromagnetic emissions before large earthquakes. Besides, laboratory, as well as field experiments, showed that electromagnetic emissions are connected with the appearance of faults. Extremely interesting works of known scientists are dedicated to the earthquake problem, but the possibility of its prediction, that is, the possibility to determine simultaneously the time and place of its occurrence and its magnitude by a definite accuracy is not seen up to now.

Researchers are still faced with complete disclosure of ongoing physical processes in the zone of earthquake preparation, and associations between many important and interesting precursor phenomena. This process, of course, requires the creation of a harmonized holistic theory, as a grandiose scientific and practical problem awaiting its solution.

In the presented work, based on a model of electromagnetic radiation generation created on the principles of electrodynamics, the authors propose theoretical and experimental studies of a precursor capable of predicting earthquakes. There are considered earthquake exogenous trigger and indicators for the Caucasus s/a region that clearly demonstrate differences between them and earthquake precursor and show the necessity to separate them from each other.

To solve the earthquake problem globally in the shortest time, it is necessary:

> First, in order to guarantee a short-term prediction of a strong earthquake in any country, it is necessary to organize a network of mobile VLF/LF EM radiation receivers, since their mobility ensures the detection of the entire spectrum of VLF/LF EM radiation.

> Second, to carry out a deep analysis of the features and the mechanism of occurrence of all anomalous geophysical fields that existed before the earthquake, in order to distinguish the trigger, indicator, and precursor of an earthquake from each other.

> Third, it is necessary to renew work in direction of seismic risk reduction by reducing the accumulated tectonic stresses based on earthquake prediction possibility.

**Conflicts of Interest**
The authors declare no competing interests.

**Acknowledgements**
The authors are grateful to the network INFREP for providing us with the MHz and kHz fractal-electromagnetic emissions data used in this paper.